\documentclass[twocolumn,amsmath,amssymb]{revtex4}
\usepackage{graphicx}
\usepackage{dcolumn}
\usepackage{bm}
\begin{document}

\title{Controllable Continuous evolution of electronic states
in a single quantum ring }

\author{Tapash Chakraborty$^1$\thanks{Tapash.Chakraborty@umanitoba.ca},
Aram Manaselyan$^2$, Manuk Barseghyan$^2$ and David Laroze$^{3,4}$}

\affiliation{$^1$ Department of Physics and Astronomy, University of Manitoba,
Winnipeg, Canada R3T 2N2}
\affiliation{$^2$ Department of Solid State Physics, Yerevan State University, Yerevan, Armenia }
\affiliation{$^3$ Instituto de Alta Investigaci\'{o}n, Universidad de Tarapac\'{a},
Casilla 7D, Arica, Chile}
\affiliation{$^4$ Yachay Tech University, School of Physical Sciences and Nanotechnology, 
00119-Urcuqu\'{i}, Ecuador}

\affiliation{E-mail: Tapash.Chakraborty@umanitoba.ca, amanasel@ysu.am,
mbarsegh@ysu.am, dlarozen@uta.cl}

\begin{abstract}

Intense terahertz laser field is shown to have a profound effect on the
electronic and optical properties of quantum rings, where the isotropic and anisotropic
quantum rings can now be treated on equal footing. We have demonstrated that in isotropic 
quantum rings the laser field creates irregular AB oscillations that are usually 
expected in anisotropic rings. Further, we have shown for the first time that intense 
laser fields can restore the {\it isotropic} physical properties in anisotropic 
quantum rings. In principle, all types of anisotropies (structural, effective masses,
defects, etc.) can evolve as in isotropic rings, in our present approach. Most
importantly, we have found a continuous evolution of the energy spectra and
intraband optical characteristics of structurally anisotropic quantum rings to 
those of isotropic rings, in a controlled manner, with the help of a laser field.

\end{abstract}

\maketitle

Research on the electronic and optical properties of quantum confined nanoscale
structures, such as quantum dots and quantum rings has made great strides in
recent years in unraveling new phenomena and their enormous potentials in device
applications. In this context, quantum rings with its doubly-connected structure
attracts special attention. Its unique topolocal structure provides a rich variety
of fascinating physical phenomena in this system. Observation of the Aharonov-Bohm
(AB) oscillations \cite{Aharonov} and the persistent current \cite{Buttiker} in small
semiconductor quantum rings (QR), and recent experimental realization of QRs with only
a few electrons \cite{Lorke,haug} have made QRs an attractive topic of experimental
and theoretical studies for various quantum effects in these quasi-one-dimensional
systems \cite{Chakraborty}. In particular, recent work has indicated the great
potentials of QRs as basis elements for a broad spectrum of applications, starting with
terahertz detectors \cite{Huang}, efficient solar cells \cite{Wu1} and memory devices
\cite{Young}, through electrically tunable optical valves and single photon emitters
\cite{Warburton,Abbarchi}. We have also worked previously on QRs in new materials
such as graphene systems \cite{graphene} and ZnO \cite{ZnO} with interesting outcomes
reported in \cite{abergel} and \cite{aram_2016} respectively.

Although almost circular or slightly oval shaped QRs have been fabricated by various
experimental groups \cite{Fominbook,Blossey,Offermans,Kleemans}, anisotropic QRs are the
ones most commonly obtained during the growth process \cite{Offermans,Wu,Raz,Sormunen}.
Theoretically the effect of anisotropy on electronic, magnetic and optical properties of
quantum rings have been investigated by various authors \cite{Planeles,Farias,Peeters2008,%
Majorana,Sousa}. In those studies, different types of anistropies were explored. For
example, in Ref.~\cite{Planeles,Peeters2008,Majorana} the shape anisotropy of the QR was
considered, while in Ref.~\cite{Farias} the anisotropy associated with defects was studied,
and in \cite{Sousa} the effective mass anisotropy was investigated. In all these cases it
was shown that the anisotropy can dramatically alter the AB oscillations in the QR. In
particular, in Ref.~\cite{Sousa} it was demonstrated that the unusual AB oscillations
caused by the effective mass anisotropy in the QR can be converted to usual AB
oscillations if the QR has an elliptical shape. However, in order to experimentally
confirm these results, one needs to grow QRs with different anisotropies and compare
their measurable optical and magnetic characteristics individually.

Here we consider the effect of a terahertz intense laser field (ILF) on isotropic and
anisotropic QRs and demonstrate that in the case of isotropic QRs the ILF can create
irregular AB oscillations that are characteristics of anisotropic rings. Additionally,
we have shown for the first time that in case of anisotropic QRs the ILF can be used as
an anisotropy controlling tool with the help of which it will be possible to visualize
both the isotropic and anisotropic properties on a single QR. For example, we have
shown that the irregular AB oscillations obtained for the elliptic QR can be made
regular with the help of the ILF. Therefore the ILF can unify all the electronic properties
of isotropic and anisotropic rings in a single system.

Our system consists of a two-dimensional anisotropic QR structure containing
electrons that are under the action of laser radiation and an external magnetic field
that is oriented along the growth direction. The laser field is represented by a monochromatic
plane wave of frequency $\omega$. The laser beam is non-resonant with the semiconductor
structure, and linearly polarized along a radial direction of the structure (chosen along
the $x$- axis). In the non-interacting case, the electron motion is described by the solution
of the time dependent Schr\"{o}dinger equation
\begin{align}
\left[\frac1{2m}\left(\widehat{\textbf{p}}-\frac{e}{c}(\textbf{A}(t)+\textbf{A}_{m})
\right)^{2}+V(x,y)\right]\Phi(x,y,t) \nonumber\\
 =i\hbar\frac{\partial}{\partial t}\Phi(x,y,t)\,,
\end{align}
where $m$ is the electron effective mass, $\textbf{p}$ is the lateral momentum of the electron,
$e$ is the absolute value of the electron charge, $\textbf{A}(t)=\textbf{e}^{}_xA^{}_0\cos(
\omega t)$ is the laser field's vector potential, where $\textbf{e}^{}_x$ denotes the unit vector
on the $x$-axis. In Eq.~(1), $\textbf{A}^{}_m$ is the vector potential of the magnetic field
which is chosen to be $\textbf{A}^{}_m=(0,Bx,0)$. In this case the scalar product
$(\textbf{A}(t)\cdot\textbf{A}^{}_m)=0$.

For the lateral confinement potential $V(x,y)$ we have chosen the model of finite, square-well
type, which can be written as:
\begin{equation}
V(x,y) = \left\{ \begin{array}{ll}
      0, & \mbox{if}\, R^{}_1\leq \sqrt{x^2+(y/\sqrt{1-\varepsilon^2})^{2}} \leq R^{}_2,\\
      V^{}_0, & {\rm otherwise},\end{array}\right.
\end{equation}
where $R^{}_1$ and $R^{}_2$ are the inner and outer radii of the QR respectively, the $\varepsilon$
describes the anisotropy of the QR ($\varepsilon=0$ corresponds to the case of circular QR) \cite{Sousa}.

Using the dipole approximation and Kramers-Hennerberger unitary transformation \cite{Kramers}
in the high-frequency regime \cite{Gavrila1,Pont,Valadares,Gavrila2,Ganichev} the laser-dressed
energies of the QR can be obtained from the following time-independent Schr\"{o}dinger equation
\begin{align}
\left[\frac1{2m}\left(\widehat{\textbf{p}}-\frac{e}{c}\textbf{A}^{}_m\right)^2+V^{}_d(x,y)
\right]\Phi^{}_d(x,y) \nonumber\\
     =E^{}_d\Phi^{}_d(x,y)\,,
\end{align}
where $V^{}_d(x,y)$ is the time-averaged laser-dressed potential that can be expressed by
the following analytical expression \cite{Radu1,Radu2}
\begin{align}
V^{}_d(x,y)&=\frac{V^{}_0}{\pi}Re\biggl[\pi- \theta\left(\alpha_{0}-x-\Gamma^{}_1
\right)\arccos\left(\frac{\Gamma^{}_1+x}{\alpha^{}_0}\right) + \nonumber\\
           & + \theta\left(\alpha^{}_0-x -\Gamma^{}_2\right)\arccos\left(
\frac{\Gamma^{}_2+x }{\alpha^{}_0}\right)-\nonumber\\
           & - \theta\left(\alpha^{}_0+x -\Gamma^{}_1\right)\arccos\left(\frac{
\Gamma^{}_1-x }{\alpha^{}_0}\right)+\nonumber\\
           & + \theta\left(\alpha^{}_0+x -\Gamma^{}_2\right)\arccos\left(\frac{
\Gamma^{}_2-x }{\alpha^{}_0}\right)           \biggr] \,,
\end{align}
where $\theta(u)$ is the Heaviside unit-step function and $\Gamma^{}_i=Re(\sqrt{R_i^2-(y/
\sqrt{1-\varepsilon^2})^2})$. The parameter $\alpha^{}_0=(e/m\epsilon_h^{1/4}\nu^2)
\sqrt{I/(2c\pi^3)}$ describes the strength of the laser field, and comprises both the intensity
$I$ and frequency $\nu$ of the laser field that can be chosen for a broad range in units of
KW/cm$^2$ and THz correspondingly \cite{Ganichev}. $\epsilon^{}_h$ is the high-frequency
dielectric constant of the system. In Fig.~1 the schematic picture of dressed potential
for different values of the ILF parameter $\alpha^{}_0$ is presented. The circular and
elliptic cases of QRs are shown. The laser dressed energy eigenvalues $E^{}_d$ and eigenfunctions
$\Phi^{}_d(x,y)$ may be obtained by solving Eq.~(3) with the help of the exact diagonalization
technique. The eigenfunctions are presented as a linear expansion of the eigenfunctions of
the two-dimensional rectangular infinitely high potential well \cite{Radu1,Radu2}.

\vspace{0.3cm}
\begin{figure}
    \centering
 \resizebox{0.9\columnwidth}{!}{\includegraphics[width=0.35\textwidth,angle=0]{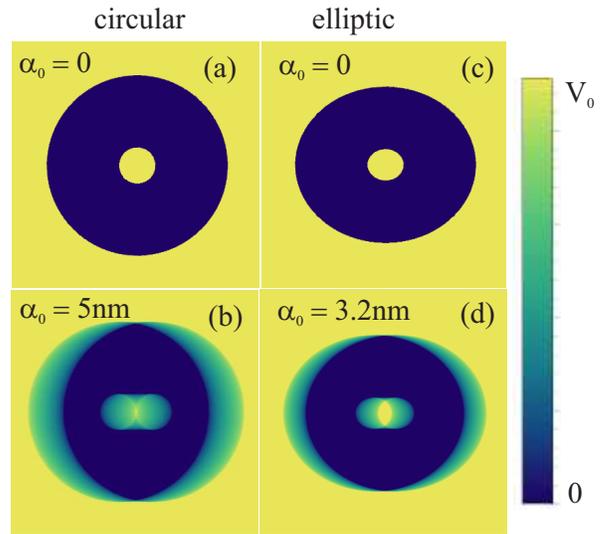}}
    \caption{The density plot of the dressed confinement potential for different values of
the ILF parameter $\alpha^{}_0$ for both circular and elliptic QRs.}
\end{figure}
\vspace{0.3cm}

We have also considered here the intraband optical transitions in the conduction band. According
to the Fermi golden rule for the $x$-polarization of the incident light the intensity of absorption
in the dipole approximation is proportional to the square of the matrix element $M_{fi}=\langle
f|x|i\rangle$, when the transition goes from the initial state $|i\rangle$ to the final state
$|f\rangle$. In this work we always consider $|i\rangle$ to be the ground state.

\vspace{0.3cm}
\begin{figure}
    \centering
 \resizebox{0.99\columnwidth}{!}{\includegraphics[width=0.35\textwidth,angle=0]{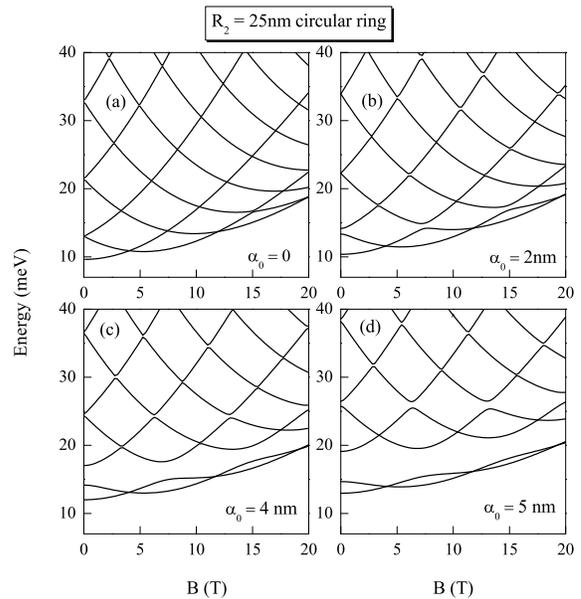}}
    \caption{The low-lying energy levels of a circular QR as a function of the magnetic field
$B$ for different values of the laser field parameter $\alpha^{}_0$. The results are for
$R^{}_2=25$ nm.}
\end{figure}
\vspace{0.3cm}

Our numerical studies are carried out for $GaAs$ QRs having parameters $V^{}_0=228$meV,
$m=0.067m^{}_0$ ($m^{}_0$ is the free electron mass), $\epsilon^{}_h=10.5$, $R^{}_1=5$ nm
\cite{Adachi}. In Fig.~2, the low-lying energy levels of circular QR with outer radius
$R^{}_2=25$ nm are presented as a function of the magnetic field $B$ for various values of
the laser field parameter $\alpha^{}_0$. In Fig.~2 (a) the regular AB effect has been
observed without the laser field, which corresponds to the case of a circular QR. The ILF
applied on a QR creates an anisotropy in the confinement potential [Fig.~1 (b)] as a result
of which the effective length of the confinement along the $x$-direction decreases in the
lower part of QR potential well. It is worth noting that with the increase of $\alpha^{}_0$,
the anisotropy of the QR is strengthened and the degeneracy of the excited states at $B=0$
disappears. With an increase of $\alpha^{}_0$ due to the reducing symmetry from $C^{}_{\infty}$
to $C^{}_2$, one should expect an energy spectrum split into non-crossing pairs of states
which in turn cross repeatedly as $B$ increases (each pair of repeatedly crossing states
containing one instance of each of two $C^{}_2$ symmetries). A similar behavior of the energy
levels, which can be called `unusual' or `irregular' AB oscillations, have been reported
earlier in QRs by other authors that is caused by the effective mass anisotropy \cite{Sousa,Peeters2008}
and structural distortions in QRs \cite{Planeles}. For $\alpha^{}_0=2$ nm, only the ground
and first excited states feel the deformation of the potential (see Fig.~2(b)). Whereas, for
larger values of $\alpha^{}_0$ more excited states start to feel the deformation of the QR
confinement potential and the two periodically crossing pairs can be visible [Fig.2~(d)].

\vspace{0.3cm}
\begin{figure}
    \centering
 \resizebox{0.99\columnwidth}{!}{\includegraphics[width=0.35\textwidth,angle=0]{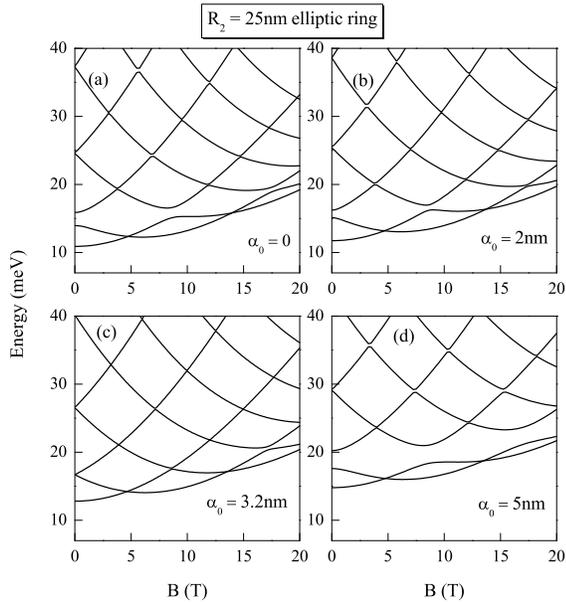}}
    \caption{The low-lying energy levels of an elliptic QR as a function of the
magnetic field $B$ for different values of $\alpha^{}_0$. The results are for $R^{}_2=25$ nm.}
\end{figure}
\vspace{0.3cm}

Let us now consider the case of the anisotropic QR under the action of ILF. From Eq.~(2)
and Fig.1(c) it is clear that if $\varepsilon\neq0$ the undressed confinement potential is
anisotropic, and the QR is compressed along the $y$ direction. On the other hand the laser field
brings an anisotropy of the confinement potential along the $x$ direction. Therefore, we
have two competing different effects, the first one is caused by the structural anisotropy
of the system while the other is caused by the external ILF. In Fig.~3 the magnetic
field dependence of the low-lying energy levels are presented, for an anisotropic QR
($\varepsilon=0.5$) and for different values of $\alpha^{}_0$. Fig.~3(a) displays the irregular
AB oscillations without the ILF due to the structural anisotropy of the QR. With an increase
of $\alpha^{}_0$ the effect of structural anisotropy on the energy levels weakens (Fig.~3(b)),
and for $\alpha^{}_0=3.2$ nm the regular AB oscillations are completely recovered (Fig.~3(c)).
A farther increase of the ILF parameter again creates an anisotropy in the $x$ direction, and
again the irregular AB oscillations can be observed in Fig.~3(d). This reentrant behavior
of the electronic states have never been reported earlier in a quantum ring. Similar effects
also can be observed for smaller QRs, which is presented in Fig.~4. The influence of ILF on
the confinement potential of the QR is stronger for smaller QRs, and therefore the regular
AB oscillations are recovered for $\alpha^{}_0=2.1$ nm (see Fig.~4(c)).

\vspace{0.3cm}
\begin{figure}
    \centering
 \resizebox{0.99\columnwidth}{!}{\includegraphics[width=0.35\textwidth,angle=0]{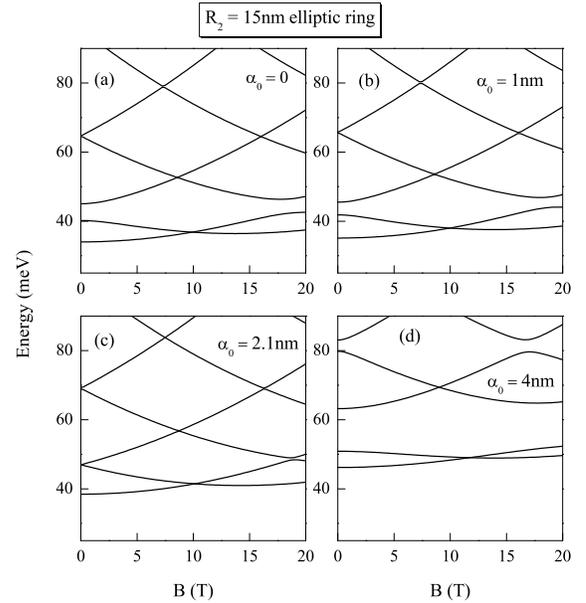}}
    \caption{The low-lying energy levels of an elliptic QR as a function of the magnetic
field $B$ for different values of $\alpha^{}_0$. The results are for $R^{}_2=15$ nm.}
\end{figure}
\vspace{0.3cm}

These interesting properties of the energy spectra are expected to influence the optical
properties of the QRs. In Fig.~5 the dipole-allowed optical transition energies as a function
of the magnetic field are presented for different values of $\alpha^{}_0$ for isotropic QR
with outer radius $R^{}_2=25$ nm. The size and the color of the circles in this figure is
proportional to the intensity of the calculated optical transitions. Without the laser field
the signature of the usual AB optical oscillations is seen in Fig.~5(a). The energies in
Fig.~5(a) correspond to the transitions from the ground state to the first and second excited
states. All other transitions are forbidden due to the cylindrical symmetry of the structure.
With the increase of $\alpha^{}_0$ the irregular optical AB oscillations are again visible.
Further, it should be noted that with the increase of $\alpha^{}_0$ the intensity of the
$1\rightarrow2$ transition weakens and the intensity of the $1\rightarrow3$ transition strengthens.
As an example, for $\alpha^{}_0=5$ nm the $1\rightarrow2$ transition has almost disappeared.
This fact can be explained by the anisotropy of the system created by the ILF in the $x$ direction.

\vspace{0.3cm}
\begin{figure}
    \centering
 \resizebox{0.99\columnwidth}{!}{\includegraphics[width=0.35\textwidth,angle=0]{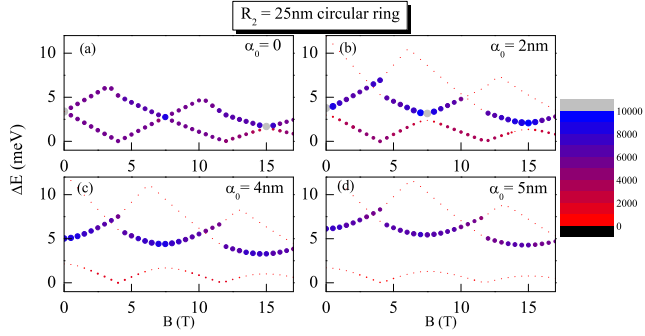}}
    \caption{Dipole allowed optical transition energies as a function of magnetic field $B$
for different values of $\alpha^{}_0$.The results are for a $R^{}_2=25$ nm circular QR. The
size and the color of the circles is proportional to the intensity of the calculated optical
transitions.}
\end{figure}
\vspace{0.3cm}

In Fig.~6 the same results as in Fig.~5 are presented for an anisotropic QR and for the value
$\varepsilon=0.5$. Without the laser field the optical AB oscillations again have irregular
behavior [Fig.~6(a)], but now the intensity of the $1\rightarrow2$ transition is
stronger than that of $1\rightarrow3$. This is because the structural anisotropy is created in
the $y$ direction. With an increase of $\alpha^{}_0$ the intensity of $1\rightarrow3$
increases and the intensity of $1\rightarrow2$ decreases. For the value of $\alpha^{}_0=3.2$ nm
the regular optical AB oscillations are completely recovered for an anisotropic QR (Fig.~6(c)).
Therefore we believe that these interesting effects can be confirmed experimentally.

\vspace{0.3cm}
\begin{figure}
    \centering
 \resizebox{0.99\columnwidth}{!}{\includegraphics[width=0.35\textwidth,angle=0]{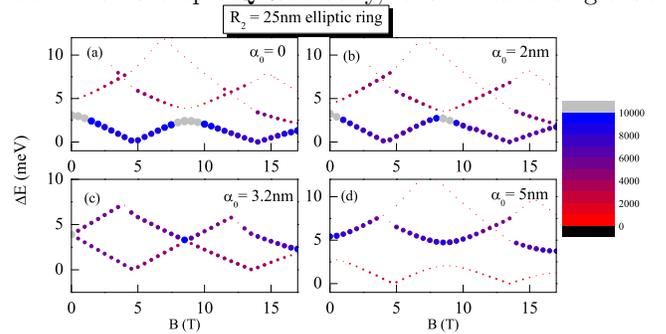}}
    \caption{Dipole-allowed optical transition energies as a finction of the magnetic field $B$
for different values of the laser field parameter $\alpha^{}_0$.The results are for $R^{}_2=25$
nm elliptic QR. The size and the color of the circles in the figure are proportional to the
intensity of the calculated optical transitions.}
\end{figure}
\vspace{0.3cm}

In conclusion, we have studied here the strong influence of intense terahertz laser field on the
electronic and optical properties of isotropic and anisotropic QRs in an applied magnetic field.
We have shown that in isotropic QRs the laser field creates the irregular AB oscillations, which
is usually expected in anisotropic rings. Therefore with the laser field we can observe
a continues evolution of AB oscillations within the same ring. In the case of anisotropic QRs we
have shown that with the ILF it is possible to completely `regularize' the anisotropy of the QR
and thus also the physical characteristics. In particular we have shown here that energy spectra
and AB oscillations have been made completely regular by the ILF for anisotropic QRs. Lastly, it
is worth noting that the ILF can in principle restore isotropic properties of a QR from any type
of anisotropy (structural, effective masses, defects, etc.) of the QRs. We believe that in addition
to providing an unified picture of the electronic and optical properties of quantum rings, our
studies will also open up new possibilities to design, fabricate and improve new devices based
on QRs, such as therahertz detectors, efficient solar cells, photon emitters, etc.

The work has been supported by the Canada Research Chairs Program of the Government of Canada,
the Armenian State Committee of Science (Project no. 15T-1C331), CONICYT-ANILLO ACT 1410, and
the center of excellence with BASAL/CONICYT financing, Grant No. FB0807, CEDENNA.

\end{document}